\documentclass[prb,showpacs,twocolumn]{revtex4}
\usepackage{epsfig}
\usepackage{graphics}
\usepackage{graphicx}
\usepackage{dcolumn}
\usepackage{bm}

\newcommand{\sba}{\begin{subeqnarray}}
\newcommand{\sea}{\end{subeqnarray}}

\def\cm-1{cm$^{-1}$}
\begin{document}

\title{Diagrammatic analysis of the Hubbard model II:\\
Superconducting state}
\author{V.\ A.\ Moskalenko$^{1,2}$}
\email{mosckalen@theor.jinr.ru}
\author{L.\ A.\ Dohotaru$^{3}$}
\author{D.\ F.\ Digor$^{1}$}
\author{I.\ D.\ Cebotari$^{1}$}
\affiliation{$^{1}$Institute of Applied Physics, Moldova Academy of Sciences, Chisinau
2028, Moldova}
\affiliation{$^{2}$BLTP, Joint Institute for Nuclear Research, 141980 Dubna, Russia}
\affiliation{$^{3}$Technical University, Chisinau 2004, Moldova}
\date{\today}

\begin{abstract}
%
Diagrammatic analysis for normal state of Hubbard model proposed in our
previous paper$^{[1]}$ is generalized and used to investigate
superconducting state of this model. We use the notion of charge quantum
number to describe the irreducible Green's function of the superconducting
state. As in the previous paper we introduce the notion of tunneling Green's
function and of its mass operator. This last quantity turns out to be equal
to correlation function of the system. We proved the existence of exact
relation between renormalized one-particle propagator and thermodynamic
potential which includes integration over auxiliary interaction constant.
The notion of skeleton diagrams of propagator and vacuum kinds were
introduced. These diagrams are constructed from irreducible Green's
functions and tunneling lines. Identity of this functional to the
thermodynamic potential has been proved and the stationarity with respect to
variation of the mass operator has been demonstrated. %
\end{abstract}

\pacs{71.27.+a, 71.10.Fd}
\maketitle

\section{Introduction}

The present paper generalizes our previous work$^{[1]}$ on diagrammatic
analysis of the normal state of the Hubbard model$^{[2-4]}$ to the
superconducting state.

Now we shall assume the existence of pairing of charge carriers and non-zero
Bogolyubov quasi-averages$^{[5]}$ and, consequently, of the Gor'kov
anomalous Green's functions$^{[6]}$.

The main property of the Hubbard model consists in the existence of strong
electron correlations and, as a result, of the new diagrammatic elements
with the structure of Kubo cumulants and named by us as irreducible Green's
functions. These functions describe the main charge, spin and pairing
fluctuations of the system.

The new diagram technique for such strongly correlated systems has been
developed in our earlier papers${[7-17]}$. This diagram technique uses the
algebra of Fermi operators and relies on the generalized Wick theorem which
contains, apart from usual Feynman contributions, additional irreducible
structures. These structures are the main elements of the diagrams.

In superconducting state, unlike the normal one, the irreducible Green's
functions can contain any even number of fermion creation and annihilation
operators, whereas in normal state the number of both kinds is equal.
Therefore we need an automatic mathematical mechanism which takes into
account all the possibilities to consider the interference of the particles
and holes in the superconducting state.

With this purpose we use the notion of charge quantum number, introduced by
us in ${[7]}$ and called $\alpha -$number, which has two values $\alpha =\pm
1$ according to the definition
\begin{eqnarray}
C^{\alpha}=\left\{\begin{array}{c} C\quad,\quad \alpha=1; \\
C^{+},\quad \alpha=-1. \end{array} \right .\label{1}
\end{eqnarray}
Were $C$ is a Fermion annihilation operator. In this new
representation the
tunneling part of the Hubbard Hamiltonian can be rewritten in the form %
\begin{eqnarray}
H^{\prime } &=&\sum\limits_{\sigma }\sum\limits_{\overrightarrow{x}%
\overrightarrow{x}^{\prime }}t(\overrightarrow{x}^{\prime }-\overrightarrow{x%
})C_{\overrightarrow{x}^{\prime }\sigma }^{+}C_{\overrightarrow{x}\sigma } \\
&=&\frac{1}{2}\sum\limits_{\alpha =-1,1}\sum\limits_{\sigma }\sum\limits_{%
\overrightarrow{x}\overrightarrow{x}^{\prime }}\alpha t_{\alpha }(%
\overrightarrow{x}^{\prime }-\overrightarrow{x})C_{\overrightarrow{x}%
^{\prime }\sigma }^{-\alpha }C_{\overrightarrow{x}\sigma }^{\alpha },
\nonumber  \label{2}
\end{eqnarray}%
%
with the definition of the tunneling matrix elements %
\begin{eqnarray}
t_{1}(\overrightarrow{x}^{\prime }-\overrightarrow{x}) &=&t(\overrightarrow{x%
}^{\prime }-\overrightarrow{x})  \nonumber \\
t_{-1}(\overrightarrow{x}^{\prime }-\overrightarrow{x}) &=&t(\overrightarrow{%
x}-\overrightarrow{x}^{\prime }) \\
t(\overrightarrow{x}=0) &=&0.  \nonumber  \label{3}
\end{eqnarray}%
%
In this charge quantum number representation the operator $H^{\prime }$ has
an additional multiple $\alpha $ for every vertex of the diagrams and
additional summation over $\alpha $. All the Green's functions depend of
this number.

In interaction representation operator $H^{\prime}$ has a form %
\begin{eqnarray}
H^{\prime}(\tau) &=&\frac{1}{2}\sum\limits_{\alpha\sigma}\sum\limits_{%
\overrightarrow{x}\overrightarrow{x}^{\prime}}\alpha t_{\alpha}(%
\overrightarrow{x}^{\prime}-\overrightarrow{x}) \\
&\times& C_{\overrightarrow{x}^{\prime}\sigma }^{-\alpha}(\tau+\alpha0^{+})
C_{\overrightarrow{x}\sigma }^{\alpha}(\tau),  \nonumber  \label{4}
\end{eqnarray}

The main part of the Hubbard Hamiltonian %
\begin{eqnarray}
H &=&H^{0}+H^{\prime },  \nonumber \\
H^{0} &=&\sum\limits_{i}H_{i}^{0}, \\
H_{i}^{0} &=&-\mu \sum\limits_{\sigma }C_{i\sigma }^{+}C_{i\sigma
}+Un_{i\uparrow }n_{i\downarrow },  \nonumber  \label{5}
\end{eqnarray}%
%
contains the local part $H^{0}$, where $\mu $ is the chemical potential and $%
U$ is the Coulomb repulsion of the electrons. This interaction is considered
as a main parameter of the model and is taken into account in zero
approximation of our theory. The operator $H^{\prime }$ describes electron
hopping between lattice sites of the crystal and is considered as a
perturbation.

We shall use the grand canonical partition function in our thermodynamic
perturbation theory.

The paper is organized in the following way. In section II we determine the
definition of one-particle Matsubara Green's functions by using $\alpha $
representation and develop the diagrammatic theory in the strong coupling
limit.

In section III we establish relation between the full thermodynamic
potential and the renormalized one-particle Green's function in the presence
of additional integration over auxiliary constant of interaction $\lambda $
and prove the stationarity theorem both for a special functional consisting
of skeleton diagrams and for a renormalized thermodynamic potential shown to
be its equivalent. 

\section{Diagrammatic theory}

%
We shall use the following definition of the Matsubara Green's functions in
the interaction representation 
\begin{eqnarray}
G^{\alpha \alpha ^{\prime }}(x|x^{\prime })=-\left\langle TC_{%
\overrightarrow{x}\sigma }^{\alpha }(\tau )C_{\overrightarrow{x}^{\prime
}\sigma ^{\prime }}^{-\alpha ^{\prime }}(\tau ^{\prime })U(\beta
)\right\rangle _{0}^{c},
\end{eqnarray}%
%
%
where $x$ stands for $(\overrightarrow{x},\sigma ,\tau )$, index $c$ of $%
\left\langle ...\right\rangle _{0}^{c}$ means the connected part of the
diagrams and $\left\langle ...\right\rangle _{0}$ means thermal average with
zero order partition function $\frac{e^{-\beta H^{0}}}{Tre^{-\beta H^{0}}}$.

We use the series expansion for the evolution operator $U(\beta)$ with some
generalization because we introduce the auxiliary constant of interaction $%
\lambda$ and use $\lambda H^{\prime}$ instead $H^{\prime}$: %
\begin{eqnarray}
U_{\lambda}(\beta ) & = & T\exp (-\lambda\int\limits_{0}^{\beta }
H^{\prime}(\tau )d\tau ),  \label{7}
\end{eqnarray}
%
with $T$ as the chronological operator. At the last stage of calculation
this constant $\lambda$ will be put equal to 1.

The correspondence between definition (6) and usual one $^{[12]}$ is the
following: 
\begin{eqnarray}
G_{\lambda }^{1,1}(x|x^{\prime }) &=&-\left\langle TC_{\overrightarrow{x}%
\sigma }(\tau )\overline{C}_{\overrightarrow{x}^{\prime }\sigma ^{\prime
}}(\tau ^{\prime })U_{\lambda }(\beta )\right\rangle _{0}^{c}  \nonumber \\
&=&G_{\sigma ,\sigma ^{\prime }}^{\lambda }(\overrightarrow{x},\tau |%
\overrightarrow{x}^{\prime },\tau ^{\prime }),  \nonumber \\
G_{\lambda }^{1,-1}(x|x^{\prime }) &=&-\left\langle TC_{\overrightarrow{x}%
\sigma }(\tau )C_{\overrightarrow{x}^{\prime }\sigma ^{\prime }}(\tau
^{\prime })U_{\lambda }(\beta )\right\rangle _{0}^{c}  \nonumber \\
&=&F_{\sigma ,\overline{\sigma }^{\prime }}^{\lambda }(\overrightarrow{x}%
,\tau |\overrightarrow{x}^{\prime },\tau ^{\prime }), \\
G_{\lambda }^{-1,1}(x|x^{\prime }) &=&-\left\langle T\overline{C}_{%
\overrightarrow{x}\sigma }(\tau )\overline{C}_{\overrightarrow{x}^{\prime
}\sigma ^{\prime }}(\tau ^{\prime })U_{\lambda }(\beta )\right\rangle
_{0}^{c}  \nonumber \\
&=&\overline{F}_{\overline{\sigma },\sigma ^{\prime }}^{\lambda }(%
\overrightarrow{x},\tau |\overrightarrow{x}^{\prime },\tau ^{\prime }),
\nonumber \\
G_{\lambda }^{-1,-1}(x|x^{\prime }) &=&-G_{\lambda }^{1,1}(x^{\prime }|x).
\nonumber  \label{8}
\end{eqnarray}%
%
%
As a result of application of the Generalized Wick Theorem we obtain for
propagator (6) the diagrammatic contributions depicted on the Fig.1 %
\begin{figure*}[t]
%
\centering
\includegraphics[width=0.85\textwidth,clip]{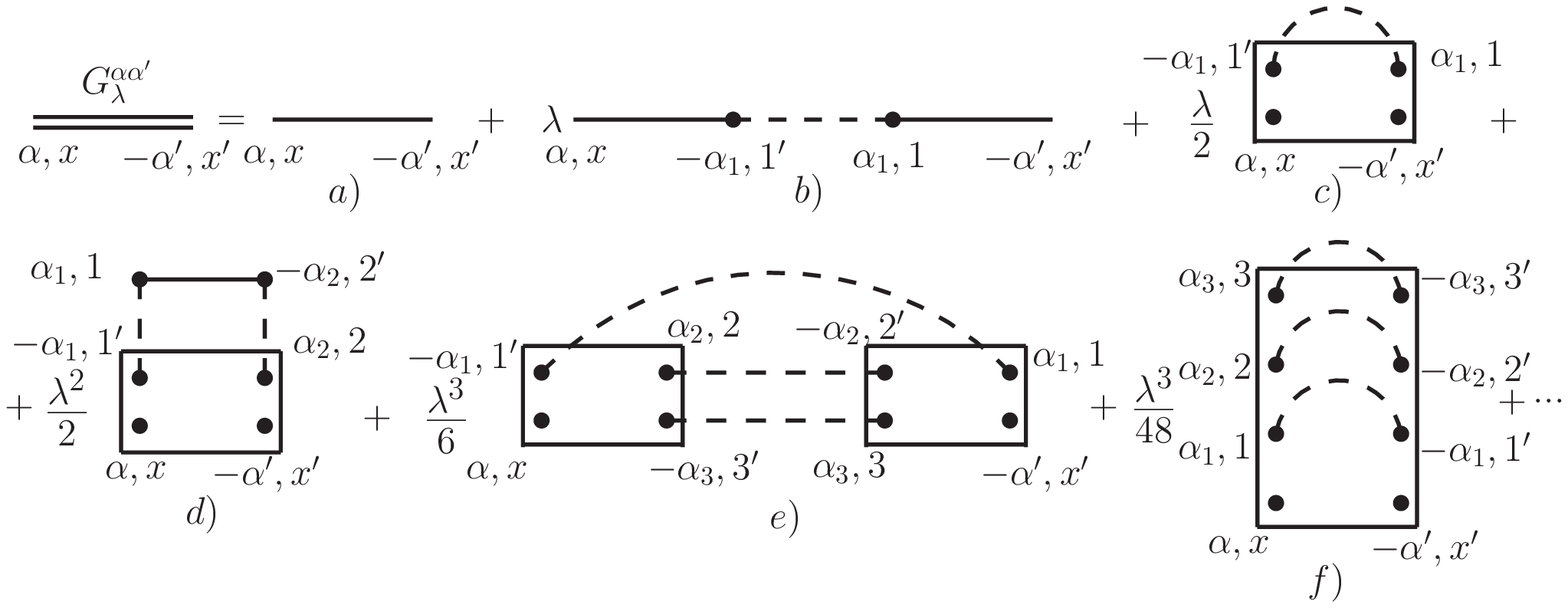} \vspace{-0mm} %
\vspace{-5mm}
\caption{The examples of the first orders perturbation theory diagrams for
propagator. Solid thin lines depict zero order one-particle Green's
functions and rectangles depict two- and four-particle irreducible Green's
functions. Thin dashed lines correspond to tunneling matrix elements. Double
solid line corresponds to renormalized propagator.}
\label{fig-1}
\end{figure*}
%
In superconducting state, unlike the normal state, the propagator lines do
not contain arrows which determine the processes of creation and
annihilation of electrons because indices $\alpha $ can take two values $%
\alpha =\pm 1$ and every vertex of the diagram describes different
possibilities.

In Fig.1 the diagram a) is the zero order propagator, the diagram b) and
more complicated diagrams of such kind are of chain type. They correspond to
the contribution of the ordinary Wick theorem and give the Hubbard I
approximation. The contributions of the diagrams c) and d) of Fig.1 are %
\begin{eqnarray}
c) &=&\frac{1}{2}\left\langle TC_{\overrightarrow{x}}^{\alpha }(\tau )C_{%
\overrightarrow{1}^{\prime }}^{-\alpha _{1}}(\tau _{1}+\alpha _{1}0^{+})C_{%
\overrightarrow{1}}^{\alpha _{1}}(\tau _{1})C_{\overrightarrow{x}^{\prime
}}^{-\alpha ^{\prime }}(\tau ^{\prime })\right\rangle _{0}^{ir}  \nonumber \\
&\times &\alpha _{1}t_{\alpha _{1}}(\overrightarrow{1}^{\prime }-%
\overrightarrow{1}),  \nonumber \\
d) &=&\frac{1}{2}\left\langle TC_{\overrightarrow{x}}^{\alpha }(\tau )C_{%
\overrightarrow{1}^{\prime }}^{-\alpha _{1}}(\tau _{1})C_{\overrightarrow{2}%
}^{\alpha _{2}}(\tau _{2})C_{\overrightarrow{x}^{\prime }}^{-\alpha ^{\prime
}}(\tau ^{\prime })\right\rangle _{0}^{ir}  \nonumber \\
&\times &\alpha _{1}t_{\alpha _{1}}(\overrightarrow{1}^{\prime }-%
\overrightarrow{1})\alpha _{2}t_{\alpha _{2}}(\overrightarrow{2}^{\prime }-%
\overrightarrow{2})  \nonumber \\
&\times &G^{(0)\alpha _{1}\alpha _{2}}(\overrightarrow{1},\tau _{1}|%
\overrightarrow{2}^{\prime },\tau _{2}),\   \nonumber
\end{eqnarray}%
%
%
where $\left\langle ...\right\rangle _{0}^{ir}$ means the irreducible
two-particle Green's function$^{[2-5]}$ and summation or integration is
understood here and below when two repeated indices are present. Spin index
has been omitted for simplicity. In the diagram c) the equality of lattice
sites indices $\overrightarrow{x}=\overrightarrow{1}^{\prime }=%
\overrightarrow{1}=\overrightarrow{x}^{\prime }$ is assumed and in diagram
d) $\overrightarrow{x}=\overrightarrow{1}^{\prime }=\overrightarrow{2}=%
\overrightarrow{x}^{\prime }$. The diagrams Fig.1 c), d) and e) contain
irreducible two-particles Green's functions, depicted as the rectangles. In
higher orders of perturbation theory more complicated many-particle
irreducible Green's functions $G_{n}^{(0)ir}[1,2,...,n]$ appear. These
functions are local, i.e. with equal lattice site indices. Therefore the
diagram c) in Fig.1 can be dropped since it contains a vanishing matrix
element, $t(\overrightarrow{x}-\overrightarrow{x})=0$. The process of
renormalization of the tunneling amplitude shown in the diagrams c) and d)
leads to the replacement of the bare tunneling matrix element $\alpha
t_{\alpha }(\overrightarrow{x}^{\prime }-\overrightarrow{x})$ in c) by a
renormalized quantity $T^{\alpha ^{\prime }\alpha }(x^{\prime }|x)$. This
process is determined by the equation %
\begin{eqnarray}
T_{\sigma ^{\prime }\sigma }^{\alpha ^{\prime }\alpha }(x^{\prime }|x)
&=&T_{\sigma ^{\prime }\sigma }^{(0)\alpha ^{\prime }\alpha }(x^{\prime
}|x)+T_{\sigma ^{\prime }\sigma _{1}}^{(0)\alpha ^{\prime }\alpha
_{1}}(x^{\prime }|x_{1}) \\
&\times &G_{\sigma _{1}\sigma _{2}}^{\alpha _{1}\alpha
_{2}}(x_{1}|x_{2})T_{\sigma _{2}\sigma }^{(0)\alpha _{2}\alpha }(x_{2}|x),
\nonumber  \label{9}
\end{eqnarray}%
%
%
where %
\begin{eqnarray}
T_{\sigma ^{\prime }\sigma }^{(0)\alpha ^{\prime }\alpha }(x^{\prime }|x)
&=&\delta _{\alpha \alpha ^{\prime }}\alpha t_{\alpha }(\overrightarrow{x}%
^{\prime }-\overrightarrow{x})\delta (\tau ^{\prime }-\tau -\alpha
0^{+})\delta _{\sigma \sigma ^{\prime }},  \nonumber  \label{10} \\
&&
\end{eqnarray}%
%
%
and $G^{\alpha _{1}\alpha _{2}}$ is the full one-particle propagator. The
quantity $T^{\alpha ^{\prime }\alpha }$ is shown in the diagrams as a double
dashed line.

We then introduce the notion of correlation function $\Lambda ^{\alpha
\alpha ^{\prime }}(x|x^{\prime })$ which is the infinite sum of strongly
connected parts of propagator's diagrams. If we now omit from these diagrams
all those contained in the process of renormalization of the tunneling
matrix element, we obtain the skeleton diagrams for correlation function. In
such skeleton diagrams we replace thin dashed lines by double dashed lines
and obtain the definition of $\Lambda ^{\alpha \alpha ^{\prime
}}(x|x^{\prime })$ shown in the Fig.2 %
\begin{figure*}[t]
%
\centering
\includegraphics[width=0.85\textwidth,clip]{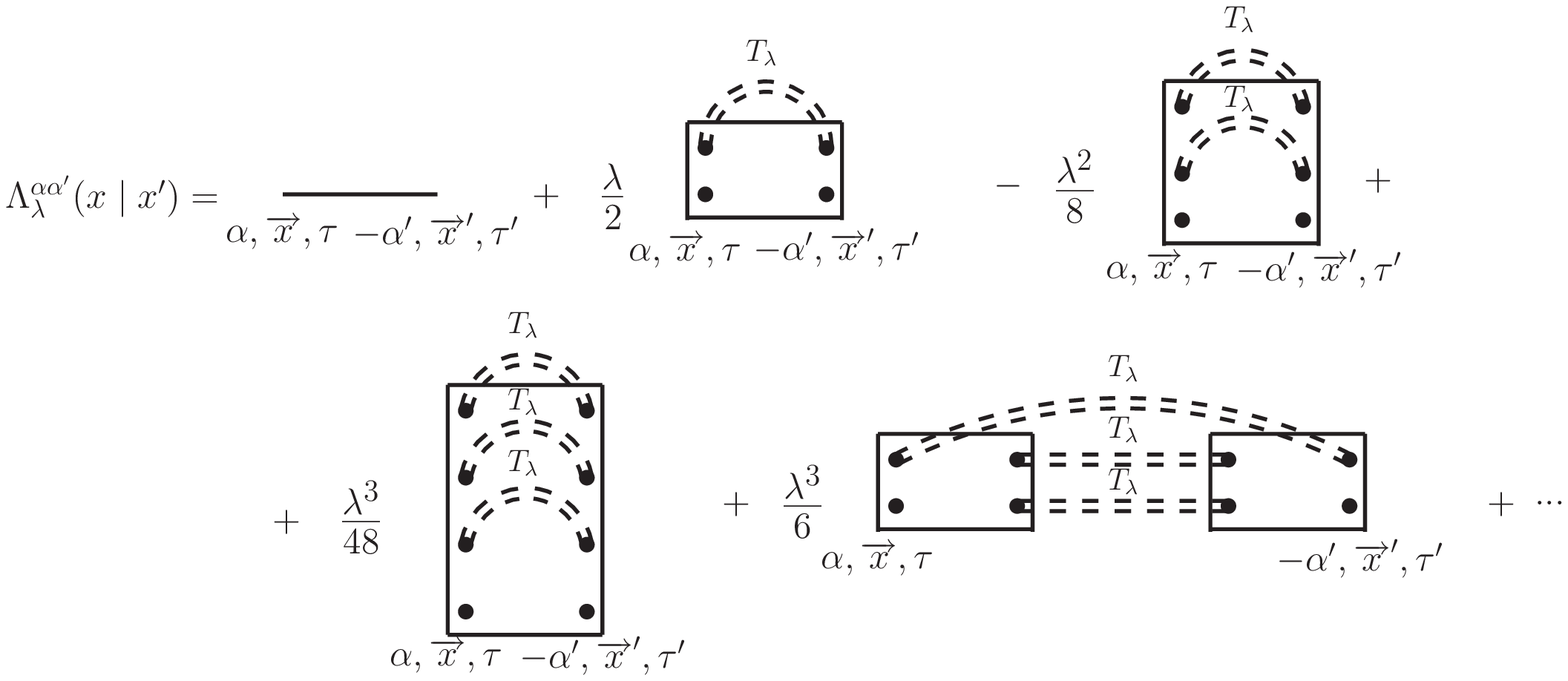} \vspace{-5mm} %
\caption{ The skeleton diagrams for correlation function $\Lambda ^{\protect%
\alpha \protect\alpha ^{\prime }}(x|x^{\prime })$. The rectangles depict the
many-particles irreducible Green's function. The double dashed lines depict
the full tunneling Green function $T_{\protect\lambda }^{\protect\alpha
\protect\alpha ^{\prime }}(x|x^{\prime }).$ }
\label{fig-2}
\end{figure*}

There are two kinds of $\lambda $ dependence in the diagrams of Fig.2. One
is conditioned by dependence of $T_{\lambda }^{\alpha \alpha ^{\prime }}$
and the second is determined by $\lambda $ being an explicit pre-factor in
the diagrams. In Hubbard I approximation only the free propagator line is
taken into account. All the contributions of Fig.2 except the last one are
local and their Fourier representation is independent of momentum. Only
these diagrams are taken into account in Dynamical Mean Field Theory$^{[18]}$%
. The last diagram of Fig.2 has the Fourier representation which depends of
momentum.

As a result of diagrammatic analysis we can formulate the Dyson-type
equation for full one-particle Green's function $(x=\overrightarrow{x},\tau)$%
: %
\begin{eqnarray}  \label{11}
G^{\alpha\alpha^{\prime}}_{\sigma\sigma^{\prime}}(x|x^{\prime})&=&\Lambda^{%
\alpha\alpha^{\prime}}_{\sigma\sigma^{\prime}}(x|x^{\prime})+
\sum\limits_{\sigma_{1}\sigma_{2}}\sum\limits_{\overrightarrow{x}_{1}%
\overrightarrow{x}_{2}}\int\limits_{0}^{\beta
}\int\limits_{0}^{\beta}d\tau_{1}d\tau_{2}  \nonumber \\
&\times&
\Lambda^{\alpha\alpha_{1}}_{\sigma\sigma_{1}}(x|x_{1})T^{(0)\alpha_{1}%
\alpha_{2}}_{\sigma_{1}\sigma_{2}}(x_{1}|x_{2})
G^{\alpha_{2}\alpha^{\prime}}_{\sigma_{2}\sigma^{\prime}}(x_{2}|x^{\prime}).
\nonumber \\
\end{eqnarray}
%

This equation can be written in the operator form: %
\begin{eqnarray}
\widehat{G}=(1-\widehat{\Lambda }\widehat{T}^{0})^{-1}\widehat{\Lambda }=%
\widehat{\Lambda }(1-\widehat{T}^{0}\widehat{\Lambda })^{-1}
\end{eqnarray}%
%

Using equations (9) and (12) we obtain the Dyson equation for the tunneling
Green's function %
\begin{eqnarray}
\widehat{T}=\widehat{T}^{0}(1-\widehat{\Lambda }\widehat{T}^{0})^{-1}=(1-%
\widehat{T}^{0}\widehat{\Lambda })^{-1}\widehat{T}^{0},
\end{eqnarray}%
%
where the correlation function $\Lambda $ has the role of mass operator for
the renormalized tunneling Green's function.

In Appendix A we demonstrate the equivalence of the equation (11) to usual$%
^{[6]}$ representation of superconducting Green's functions.


\section{Thermodynamic potential diagrams}


The thermodynamic potential of the system is determined by the connected
part of the mean value of evolution operator: %
\begin{eqnarray}
F(\lambda )=F_{0}-\frac{1}{\beta }\left\langle U_{\lambda }(\beta
)\right\rangle _{0}^{c},
\end{eqnarray}%
%
with $\lambda $ equal to 1.

In the Fig.3 are depicted the first order diagrams for $\left\langle
U_{\lambda }(\beta )\right\rangle _{0}^{c}$. %
\begin{figure*}[t]
%
\centering
\includegraphics[width=0.85\textwidth,clip]{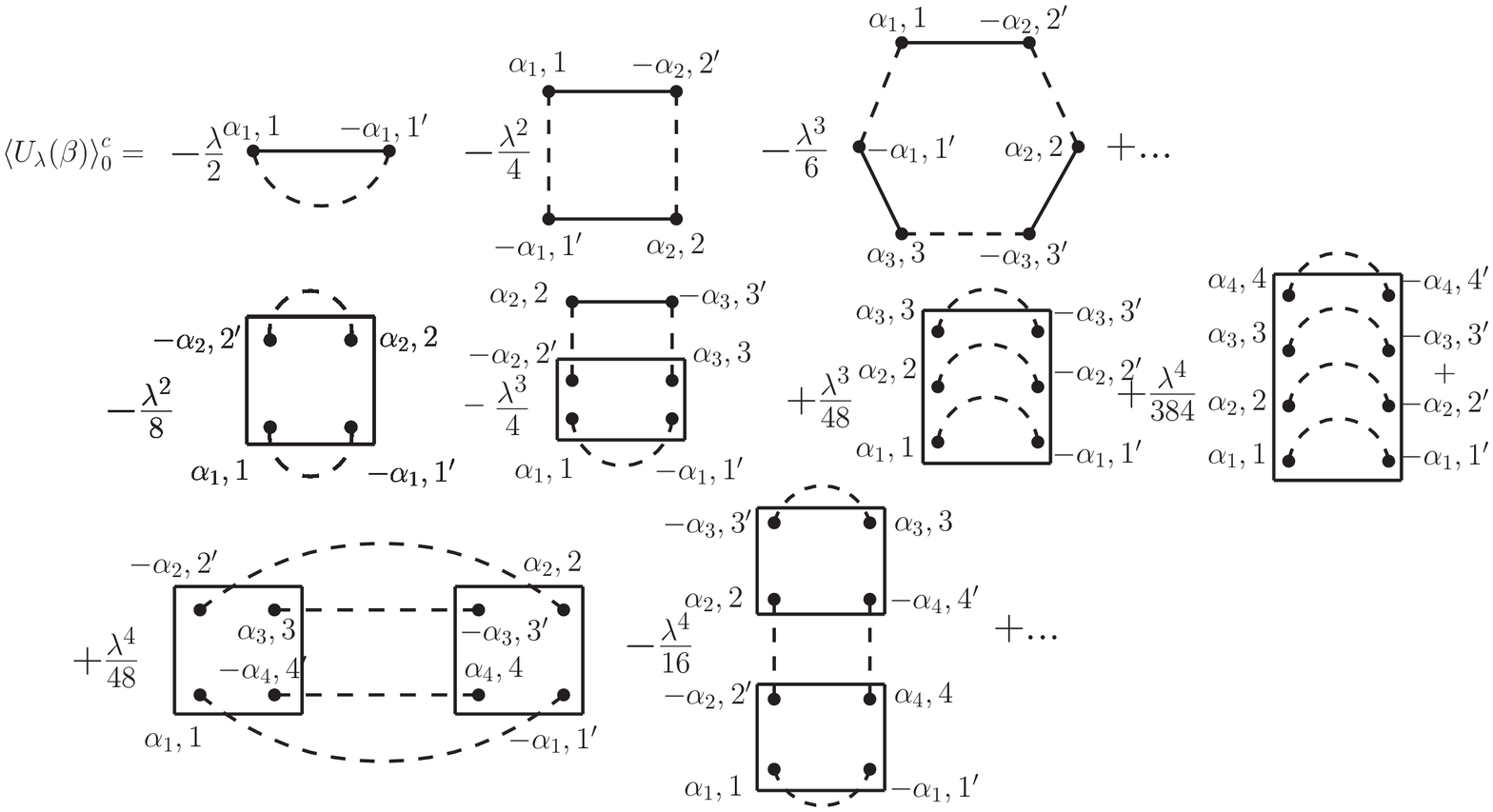} \vspace{-0mm} %
\vspace{-5mm}
\caption{ The first orders of perturbation theory contributions.}
\label{fig-3}
\end{figure*}

The notations in Fig.3 are $n=(\overrightarrow{n},\tau _{n})$ and $n^{\prime
}=(\overrightarrow{n}^{\prime },\tau _{n}+\alpha _{n}0^{+})$.

The first three diagrams in Fig.3 are of chain type and correspond
to the Hubbard I approximation. The next diagrams contain the
rectangles which represent our irreducible Green's functions.
Indeed, some of these diagrams are equal to zero when the dashed
lines are self-closed by virtue of the relation $t(0)=0$. However,
when these dashed lines are replaced by renormalized quantities
$T_{\lambda }$ their contributions are different from zero and
should be retained. Such renormalized tunneling quantities will be
used in the next part of the paper. The contributions of the fifth
and eighth diagrams on the right-hand side of Fig.3 are %
\begin{eqnarray}
&-&\frac{1}{4}\left\langle TC_{\overrightarrow{1}}^{\alpha _{1}}(\tau
_{1})C_{\overrightarrow{2}^{\prime }}^{-\alpha _{2}}(\tau _{2})C_{%
\overrightarrow{3}}^{\alpha _{3}}(\tau _{3})C_{\overrightarrow{1}^{\prime
}}^{-\alpha _{1}}(\tau _{1}+\alpha _{1}0^{+})\right\rangle _{0}^{ir}
\nonumber \\
&\times &\alpha _{1}t_{\alpha _{1}}(\overrightarrow{1}^{\prime }-%
\overrightarrow{1})\alpha _{2}t_{\alpha _{2}}(\overrightarrow{2}^{\prime }-%
\overrightarrow{2})\alpha _{3}t_{\alpha _{3}}(\overrightarrow{3}^{\prime }-%
\overrightarrow{3})  \nonumber \\
&\times &G^{(0)\alpha _{2}\alpha _{3}}(\overrightarrow{2},\tau _{2}|%
\overrightarrow{3}^{\prime },\tau _{3}),  \nonumber \\
&+&\frac{1}{48}\left\langle TC_{\overrightarrow{1}}^{\alpha _{1}}(\tau
_{1})C_{\overrightarrow{2}^{\prime }}^{-\alpha _{2}}(\tau _{2})C_{%
\overrightarrow{3}}^{\alpha _{3}}(\tau _{3})C_{\overrightarrow{4}^{\prime
}}^{-\alpha _{4}}(\tau _{4})\right\rangle _{0}^{ir}  \nonumber \\
&\times &\alpha _{1}t_{\alpha _{1}}(\overrightarrow{1}^{\prime }-%
\overrightarrow{1})\alpha _{2}t_{\alpha _{2}}(\overrightarrow{2}^{\prime }-%
\overrightarrow{2})  \nonumber \\
&\times &\left\langle TC_{\overrightarrow{4}}^{\alpha _{4}}(\tau _{4})C_{%
\overrightarrow{3}^{\prime }}^{-\alpha _{3}}(\tau _{3})C_{\overrightarrow{2}%
}^{\alpha _{2}}(\tau _{2})C_{\overrightarrow{1}^{\prime }}^{-\alpha
_{1}}(\tau _{1})\right\rangle _{0}^{ir}  \nonumber \\
&\times &\alpha _{3}t_{\alpha _{3}}(\overrightarrow{3}^{\prime }-%
\overrightarrow{3})\alpha _{4}t_{\alpha _{4}}(\overrightarrow{4}^{\prime }-%
\overrightarrow{4}),\   \nonumber
\end{eqnarray}%
%
respectively.

Comparison of the diagrams of Fig.1 for the $G^{(n)\alpha \alpha ^{\prime }}$
($n$-th order of perturbation theory for the one-particle propagator) to the
contributions of Fig.2 for $\left\langle U_{\lambda }^{n+1}(\beta
)\right\rangle _{0}^{c}$ ($(n+1)$-th order for evolution operator) allows us
to establish the following simple relation $(n\geq 1)$: %
\begin{eqnarray}
\left\langle U_{\lambda }^{(n+1)}(\beta )\right\rangle _{0}^{c} &=&-\frac{%
\beta }{2}\int\limits_{0}^{\lambda }\frac{d\lambda _{1}}{\lambda _{1}}%
\sum\limits_{\overrightarrow{1}\overrightarrow{1}^{\prime
}}\sum\limits_{\alpha _{1}\sigma _{1}}\lambda _{1}\alpha _{1}t_{\alpha _{1}}(%
\overrightarrow{1}^{\prime }-\overrightarrow{1})  \nonumber \\
&\times &G_{\sigma _{1}\lambda _{1}}^{(n)\alpha _{1}\alpha _{1}}(%
\overrightarrow{1}-\overrightarrow{1}^{\prime }|-\alpha _{1}0^{+}),
\nonumber
\end{eqnarray}%
and as the result we have %
\begin{eqnarray}
\left\langle U_{\lambda }(\beta )\right\rangle _{0}^{c} &=&-\frac{1}{2}%
\int\limits_{0}^{\lambda }\frac{d\lambda _{1}}{\lambda _{1}}\beta
\sum\limits_{\overrightarrow{1}\overrightarrow{1}^{\prime
}}\sum\limits_{\alpha _{1}\sigma _{1}}\lambda _{1}\alpha _{1}t_{\alpha _{1}}(%
\overrightarrow{1}^{\prime }-\overrightarrow{1})  \nonumber \\
&\times &G_{\sigma _{1}\lambda _{1}}^{\alpha _{1}\alpha _{1}}(%
\overrightarrow{1}-\overrightarrow{1}^{\prime }|-\alpha _{1}0^{+}) \\
&=&-\frac{1}{2}\int\limits_{0}^{\lambda }\frac{d\lambda _{1}}{\lambda _{1}}%
Tr(\lambda _{1}\widehat{T}^{0}\widehat{G}_{\lambda _{1}}).  \nonumber
\label{15}
\end{eqnarray}%

Taking into account equations (12) and (13) we obtain %
\begin{eqnarray}
\left\langle U_{\lambda }(\beta )\right\rangle _{0}^{c}=-\frac{1}{2}%
\int\limits_{0}^{\lambda }\frac{d\lambda _{1}}{\lambda _{1}}Tr(\lambda _{1}%
\widehat{T}_{\lambda _{1}}\widehat{\Lambda }_{\lambda _{1}}).
\end{eqnarray}%
Then from (14) and (16) it follows that %
\begin{eqnarray}
F(\lambda ) &=&F_{0}+\frac{1}{2\beta }\int\limits_{0}^{\lambda }\frac{%
d\lambda _{1}}{\lambda _{1}}Tr(\lambda _{1}\widehat{T}_{\lambda _{1}}%
\widehat{\Lambda }_{\lambda _{1}})  \nonumber \\
&=&F_{0}+\frac{1}{2\beta }\int\limits_{0}^{\lambda }\frac{d\lambda _{1}}{%
\lambda _{1}}Tr(\widehat{T}_{\lambda _{1}}\widehat{\Sigma }_{\lambda _{1}}),
\label{17}
\end{eqnarray}%
where %
\begin{eqnarray}
\widehat{\Sigma }_{\lambda }=\lambda \widehat{\Lambda }_{\lambda }
\end{eqnarray}%
has the role of mass operator for tunneling Green's function $\widehat{T}%
_{\lambda }$. For them Dyson equation exists: %
\begin{eqnarray}
\widehat{T}=\widehat{T}^{0}+\widehat{T}^{0}\widehat{\Sigma
}\widehat{T}.
\end{eqnarray}%
Equation (17) can be rewritten in the form %
\begin{eqnarray}
\lambda \frac{dF(\lambda )}{d\lambda }=\frac{1}{2\beta }Tr(\widehat{T}%
_{\lambda }\widehat{\Sigma }_{\lambda }).
\end{eqnarray}%
The equations (15) and (17) establish the relation between the thermodynamic
potential and renormalized one-particle propagator $\widehat{G}_{\lambda }$
or tunneling Green's function $\widehat{T}_{\lambda }$. Both these
quantities depend on auxiliary parameter $\lambda $ which is integrated
over. As have been proved by Luttinger and Ward$^{[19,20]}$, for normal
state of weakly correlated systems, it is possible to obtain another
expression for the thermodynamic potential without such additional
integration.

In our previous paper$^{[1]}$, for the normal state of Hubbard model, we
have obtained such an equation in the form of special functional. We now
consider its generalization to the case of superconductivity. For this
purpose we introduce the functional %
\begin{eqnarray}
Y(\lambda )=Y_{1}(\lambda )+Y^{\prime }(\lambda ),
\end{eqnarray}%
where %
\begin{eqnarray}
Y_{1}(\lambda )=-\frac{1}{2}Tr\{\ln (\lambda
\widehat{T}^{0}\widehat{\Lambda }_{\lambda }-1)+\widehat{T}_{\lambda
}\lambda \widehat{\Lambda }_{\lambda }\},
\end{eqnarray}%
and $Y^{\prime }(\lambda )$ is the functional constructed from skeleton
diagrams depicted on Fig.4. %
\begin{figure*}[t]
%
\centering
\includegraphics[width=0.85\textwidth,clip]{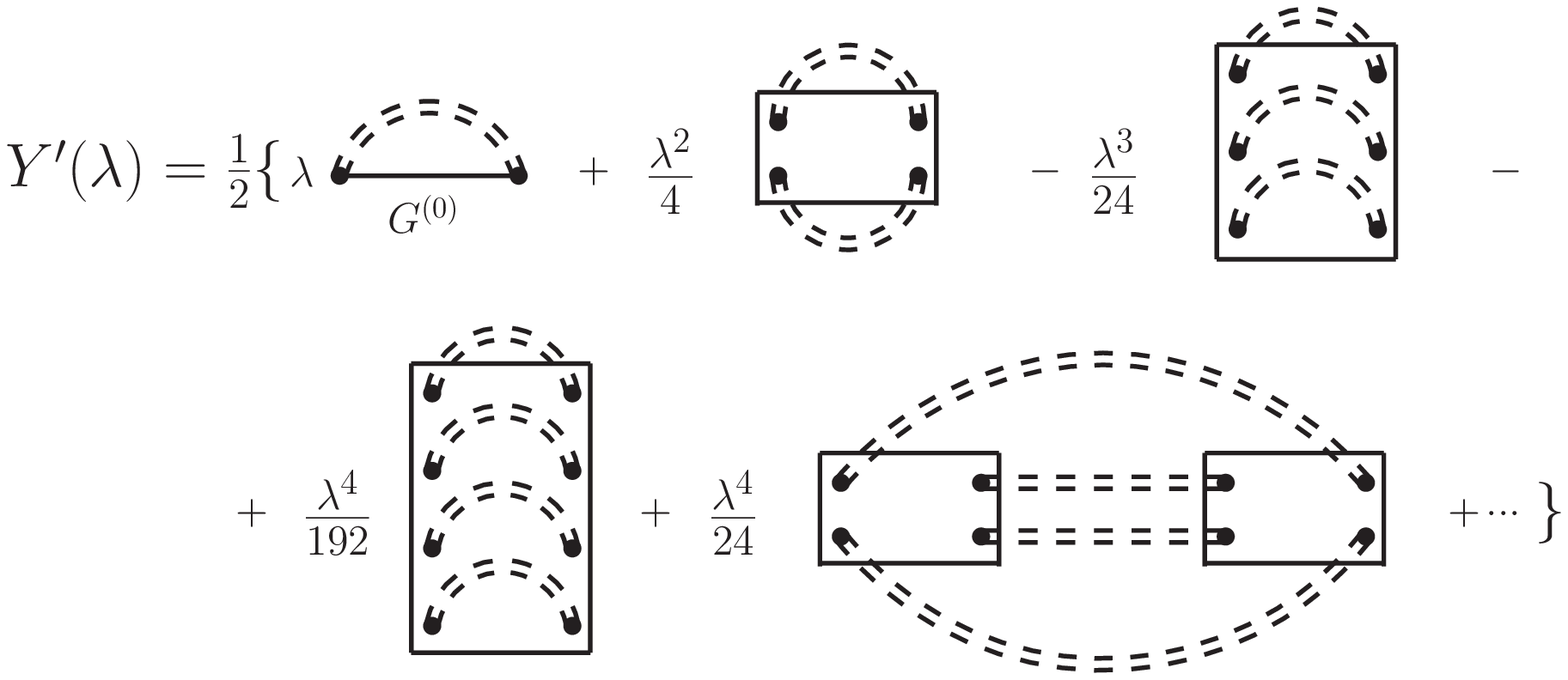} \vspace{-0mm} %
\vspace{-5mm}
\caption{ The skeleton diagrams for functional $Y^{\prime }(\protect\lambda )
$. The rectangles depict the irreducible Green's functions. The double
dashed lines depict the tunneling renormalized Green's functions $T^{\protect%
\alpha \protect\alpha ^{\prime }}(x|x^{\prime })$.}
\label{fig-4}
\end{figure*}
%

From Fig.4 and Fig.2 it is possible to obtain the relation %
\begin{eqnarray}
\frac{\delta Y^{\prime }(\lambda )}{\delta T_{\lambda }^{\alpha
\alpha ^{\prime }}(x|x^{\prime })}=\frac{1}{2}\lambda \Lambda
_{\lambda }^{\alpha ^{\prime }\alpha }(x^{\prime }|x).
\end{eqnarray}%
Now we shall take into account the following functional derivatives based on
the equation (12) and (13): %
\begin{eqnarray}
\frac{\delta }{\delta T_{\lambda }^{\alpha \alpha ^{\prime }}(x|x^{\prime })}%
Tr(\ln (\widehat{T}^{0}\widehat{\Lambda }\lambda -1)) =-Tr(\widehat{T}%
\frac{\delta \widehat{\Lambda }}{\delta T_{\lambda }^{\alpha \alpha ^{\prime
}}(x|x^{\prime }}),  \nonumber  \label{24} \\
\frac{\delta }{\delta T_{\lambda }^{\alpha \alpha ^{\prime }}(x|x^{\prime })}%
Tr(\widehat{T}\widehat{\Lambda }\lambda )=\lambda \Lambda _{\lambda
}^{\alpha ^{\prime }\alpha }(x^{\prime }|x)  \nonumber \\
+Tr(\widehat{T}_{\lambda }\frac{\delta \widehat{\Lambda }_{\lambda }}{%
\delta T_{\lambda }^{\alpha \alpha ^{\prime }}(x|x^{\prime }}\lambda ).
\nonumber \\
\end{eqnarray}%
As a consequence of these equations we have %
\begin{eqnarray}
\frac{\delta }{\delta T_{\lambda }^{\alpha \alpha ^{\prime }}(x|x^{\prime })}%
Tr\{\ln (\lambda \widehat{T}^{0}\widehat{\Lambda }_{\lambda }-1)+\lambda
\widehat{T}_{\lambda }\widehat{\Lambda }_{\lambda }\} &=&\lambda \Lambda
_{\lambda }^{\alpha ^{\prime }\alpha }(x^{\prime }|x),  \nonumber  \label{25}
\\
&&
\end{eqnarray}%
and %
\begin{eqnarray}
\frac{\delta Y_{1}(\lambda )}{\delta T_{\lambda }^{\alpha \alpha
^{\prime }}(x|x^{\prime })}=-\frac{\lambda }{2}\Lambda _{\lambda
}^{\alpha ^{\prime }\alpha }(x^{\prime }|x).
\end{eqnarray}%
With the functional derivative of $Y^{\prime }(\lambda )$ given in (23) we
obtain the stationarity property of the functional $Y(\lambda )$: %
\begin{eqnarray}
\frac{\delta Y(\lambda )}{\delta T_{\lambda }^{\alpha \alpha
^{\prime }}(x|x^{\prime })}=0
\end{eqnarray}%
%

Using the definition (18) of the mass operator $\widehat{\Sigma }_{\lambda }$
we can rewrite the functional $Y_{1}(\lambda )$ in the form %
\begin{eqnarray}
Y_{1}(\lambda )=-\frac{1}{2}Tr\{\ln (\widehat{T}^{0}\widehat{\Sigma }%
_{\lambda }-1)+\widehat{T}_{\lambda }\widehat{\Sigma }_{\lambda }\},
\end{eqnarray}%
and prove the second form of stationarity property %
\begin{eqnarray}
\frac{\delta \ Y(\lambda )}{\delta \widehat{\Sigma }_{\lambda }}=0.
\end{eqnarray}%
To demonstrate this equation it is sufficient to use the Dyson equation (19)
in the form %
\begin{eqnarray}
\widehat{{T}^{0}}^{-1}=\widehat{T}_{\lambda }^{-1}+\widehat{\Sigma }%
_{\lambda },
\end{eqnarray}%
and the derivatives: %

\begin{eqnarray}
\frac{\delta (\widehat{T}_{\lambda }^{-1})^{\beta \beta ^{\prime
}}(y|y^{\prime })}{\delta (\widehat{\Sigma }_{\lambda })^{\alpha
\alpha ^{\prime }}(x|x^{\prime })} =\delta _{\alpha \beta }\delta
_{\alpha
^{\prime }\beta ^{\prime }}\delta _{xy}\delta _{x{\prime }y{\prime }},\nonumber \\
\frac{\delta (\widehat{T}_{\lambda })^{\beta \beta ^{\prime
}}(y|y^{\prime }) }{\delta (\widehat{\Sigma }_{\lambda })^{\alpha
\alpha ^{\prime }}(x|x^{\prime })} =(\widehat{T}_{\lambda })^{\alpha
^{\prime }\beta ^{\prime }}(x^{\prime }|y^{\prime
})(\widehat{T}_{\lambda })^{\beta \alpha
}(y|x),  \nonumber \\
\frac{\delta }{\delta (\widehat{\Sigma }_{\lambda })^{\alpha \alpha
^{\prime }}(x|x^{\prime })}Tr(\widehat{T}_{\lambda }\widehat{\Sigma
}_{\lambda }) =(\widehat{T}_{\lambda })^{\alpha ^{\prime }\alpha
}(x^{\prime }|x)\\
+(\widehat{T}_{\lambda }\widehat{\Sigma }_{\lambda
}\widehat{T}_{\lambda
})^{\alpha ^{\prime }\alpha }(x^{\prime }|x),  \nonumber \\
\frac{\delta }{\delta (\widehat{\Sigma }_{\lambda })^{\alpha \alpha
^{\prime }}(x|x^{\prime })}Tr\{\ln (\widehat{T}^{0}\widehat{\Sigma
}_{\lambda }-1)\} =-(\widehat{T}_{\lambda })^{\alpha ^{\prime
}\alpha }(x^{\prime }|x). \nonumber  \label{31}
\end{eqnarray}%
Therefore we have %
\begin{eqnarray}
\frac{\delta Y_{1}(\lambda )}{\delta (\widehat{\Sigma }_{\lambda
})^{\alpha
\alpha ^{\prime }}(x|x^{\prime })}=-\frac{1}{2}(\widehat{T}_{\lambda }%
\widehat{\Sigma }_{\lambda }\widehat{T}_{\lambda })^{\alpha ^{\prime }\alpha
}(x^{\prime }|x),
\end{eqnarray}%
and %
\begin{eqnarray}
\frac{\delta Y^{\prime }(\lambda )}{\delta (\widehat{\Sigma }_{\lambda
})^{\alpha \alpha ^{\prime }}(x|x^{\prime })} &=&\frac{\delta Y^{\prime
}(\lambda )}{\delta (\widehat{T}_{\lambda })^{\beta \beta ^{\prime
}}(y|y^{\prime })}\frac{\delta (\widehat{T}_{\lambda })^{\beta \beta
^{\prime }}(y|y^{\prime })}{\delta (\widehat{\Sigma }_{\lambda })^{\alpha
\alpha ^{\prime }}(x|x^{\prime })}  \nonumber \\
&=&\frac{1}{2}(\widehat{T}_{\lambda }\widehat{\Sigma }_{\lambda }\widehat{T}%
_{\lambda })^{\alpha ^{\prime }\alpha }(x^{\prime }|x),  \label{33}
\end{eqnarray}%
where the usual convention about summation over the repeated indices has
been adopted.

As a result we obtain the stationarity property (29) of the functional $%
Y(\lambda )$ versus the change of the mass operator $\Sigma _{\lambda }$.
This mass operator for $\lambda =1$ coincides with correlation function of
our strongly correlated model.

Now it is necessary to find a relation between the thermodynamic potential $%
F(\lambda )$ and the functional $Y(\lambda )$.

Consider first the value of the derivative $\frac{dY(\lambda )}{d\lambda }$.
The $\lambda $ dependence of the functional $Y(\lambda )$ is of two kinds:
through $\Sigma _{\lambda }$ and also explicit through the factors $\lambda
^{n}$ in front of the skeleton diagrams for the functional $Y^{\prime
}(\lambda )$.

Due the stationarity property (29) we obtain %
\begin{eqnarray}
\frac{dY(\lambda )}{d\lambda } &=&\frac{\delta Y(\lambda )}{\delta \Sigma
_{\lambda }}\frac{d\Sigma _{\lambda }}{d\lambda }+\frac{\partial Y(\lambda )%
}{\partial \lambda }|_{\Sigma _{\lambda }} \\
&=&\frac{dY(\lambda )}{d\lambda }|_{\Sigma _{\lambda }}=\frac{dY^{\prime
}(\lambda )}{d\lambda }|_{\Sigma _{\lambda }}.  \nonumber  \label{34}
\end{eqnarray}%
Here we took into account that the $Y_{1}(\lambda )$ part of functional $%
Y(\lambda )$ (see equations (21) and (28)) does not explicitly dependent on $%
\lambda $.

By using the definitions of $Y^{\prime }(\lambda )$ (see Fig.4) and of $%
\Lambda _{\lambda }$ (see Fig. 2) it is easy to establish the property: %
\begin{eqnarray}
\lambda \frac{dY(\lambda )}{d\lambda } &=&\lambda \frac{\partial Y^{\prime
}(\lambda )}{\partial \lambda }|_{\Sigma _{\lambda }}=\frac{1}{2\beta }%
Tr(\lambda \widehat{T}_{\lambda }\widehat{\Lambda }_{\lambda })  \nonumber \\
&=&\frac{1}{2\beta }Tr(\widehat{T}_{\lambda }\widehat{\Sigma }_{\lambda }).
\label{35}
\end{eqnarray}%
From the equations (20) and (35) we have %
\begin{eqnarray}
\lambda \frac{dY(\lambda )}{d\lambda }=\frac{1}{2\beta }Tr(\widehat{T}%
_{\lambda }\widehat{\Sigma }_{\lambda })=\lambda \frac{dF(\lambda )}{%
d\lambda },
\end{eqnarray}%
and we therefore obtain %
\begin{eqnarray}
F(\lambda )=Y(\lambda )+F_{0},
\end{eqnarray}%
since for $\lambda =0$ the perturbation is absent $Y(\lambda =0)=0$ and $%
F(\lambda =0)=F_{0}.$ Now we set $\lambda =1$ and obtain %
\begin{eqnarray}
F=F_{0}+Y(1),
\end{eqnarray}%
with the stationarity property %
\begin{eqnarray}
\frac{\delta F}{\delta \widehat{\Sigma }}=0.
\end{eqnarray}%

\section{Conclusions}


We have further developed the diagrammatic theory proposed for strongly
correlated systems many years ago to establish the stationarity property of
the thermodynamic potential in the superconducting state of the Hubbard
model.

First, we have introduced the notion of charge quantum number which gives
the possibility to consider the presence of irreducible Green's functions
with an arbitrary number of creation or annihilation Fermi- operators in
superconducting state.

We have introduced the notion of tunneling Green's function and its mass
operator, which turns out to be equal to the correlation function of the
fermion system.

We have proven the existence of the Dyson equation for this function and
establish the exact relation between the thermodynamic potential and
renormalized one-particle propagator. This relation contains an additional
integration over the auxiliary constant of interaction $\lambda $.

We have constructed a special functional based on the skeleton diagrams for
the propagator and for the evolution operator which contain the irreducible
Green's functions and full tunneling Green's functions.

We have proven the existence of the stationarity property of this functional
and establish its relation with thermodynamic potential.

It is important to emphasize that there is a close similarity between our
results obtained for two different models of strongly correlated systems
such as Periodic Anderson Model (PAM) and the Hubbard Model (HM). From
comparison of the results obtained for the PAM (see paper [17]) and the
results of the present paper for the HM the topological coincidence of the
diagrams for both models has been revealed.

For example the skeleton diagrams of Fig. 3 of paper [17], obtained for $%
\Lambda $ functional of PAM topologically coincide with the skeleton
diagrams of our Fig. 2 for the same functional, but of quite a different
model. In order to obtain a complete coincidence, it is necessary to replace
the full Green's function $G_{c}(i\omega )$ of conduction electrons of PAM
by the renormalized tunneling Green's function $T(i\omega )$ of the HM.

The same similarity exists between other functionals of these models. For
example, comparison of the skeleton diagrams of Fig. 10 of paper [17] with
the diagrams of Fig. 4 of the present paper reveals the full coincidence
upon replacement of the Green's functions $G_{c}$ by $T$. This comparison
allows us to conclude that from the thermodynamic point of view the Periodic
Anderson Model can be reduced to the Hubbard Model if we replace the Green's
function of the conduction electrons of PAM subsystem by tunneling Green's
function of hopping electrons of HM. This equivalence is irrelevant for the
kinetic properties of PAM.

We also note that the skeleton representation of our functional allows to
select the local irreducible Green's functions as can be seen from Figs. 2
and 10. These quantities contain only fluctuations in time, unlike the non
local ones which include both fluctuations in time and space. The
coefficients of local diagrams (see Fig. 10) vary with the order of
perturbation theory as $\frac{1}{2^{n-1}n!}$ for $n>1$.

Only such local diagrams are relevant for DMFT, so that one can attempt to
carry out the summation of this class of diagrams.

%

\begin{acknowledgments}
Two of us (V.A.M. and L.A.D.) thank Professor N.M. Plakida and Doctor. S.
Cojocaru for fruitful discussion.
\end{acknowledgments}

%

%
%
\newpage
\appendix
%

\section{Gor'kov-Nambu representation}

%
We consider equation (11) in Fourier representation. By inserting specific
values of charge quantum number $\lambda =\pm 1$ we obtain $(k=(%
\overrightarrow{k},i\omega _{n}))$: %
\begin{widetext}
\begin{eqnarray}
G_{\sigma \sigma ^{\prime }}^{1,1}(k) =\Lambda _{\sigma \sigma
^{\prime }}^{1,1}(k)+\Lambda _{\sigma \sigma _{1}}^{1,1}(k)\epsilon
_{1}( \overrightarrow{k})G_{\sigma _{1}\sigma ^{\prime }}^{1,1}(k)
-\Lambda _{\sigma ,-\sigma _{1}}^{1,-1}(k)\epsilon
_{-1}(\overrightarrow{k}
)G_{-\sigma _{1},\sigma ^{\prime }}^{-1,1}(k),\label{A1} \\
G_{\sigma ,-\sigma ^{\prime }}^{1,-1}(k) =\Lambda _{\sigma ,-\sigma
^{\prime }}^{1,-1}(k)+\Lambda _{\sigma \sigma _{1}}^{1,1}(k)\epsilon
_{1}( \overrightarrow{k})G_{\sigma _{1},-\sigma ^{\prime
}}^{1,-1}(k) +\Lambda _{\sigma ,-\sigma _{1}}^{1,-1}(k)\epsilon
_{-1}(\overrightarrow{k} )G_{-\sigma ^{\prime },-\sigma
}^{1,1}(-k),\label{A2}\\
G_{-\sigma ,\sigma ^{\prime }}^{-1,1}(k) =\Lambda _{-\sigma ,\sigma
^{\prime }}^{-1,1}(k)+\Lambda _{-\sigma ,\sigma
_{1}}^{-1,1}(k)\epsilon _{1}( \overrightarrow{k})G_{\sigma
_{1}\sigma ^{\prime }}^{1,1}(k)+\Lambda _{-\sigma _{1},-\sigma
}^{1,1}(-k)\epsilon _{-1}(\overrightarrow{k
})G_{-\sigma _{1},\sigma ^{\prime }}^{-1,1}(k),\label{A3} \\
G_{-\sigma ^{\prime },-\sigma }^{1,1}(-k) =\Lambda _{-\sigma
^{\prime },-\sigma }^{1,1}(-k)-\Lambda _{-\sigma ,\sigma
_{1}}^{-1,1}(k)\epsilon _{1}( \overrightarrow{k}) G_{\sigma
_{1},-\sigma ^{\prime }}^{1,-1}(k)+\Lambda _{-\sigma _{1},-\sigma
}^{1,1}(-k)\epsilon _{-1}(\overrightarrow{k})G_{-\sigma ^{\prime
},-\sigma _{1}}^{1,1}(-k).  \label{A4}
\end{eqnarray}
\end{widetext}%
%
Here %
\begin{widetext}
\begin{eqnarray}
\epsilon _{1}(\overrightarrow{k})=\epsilon
(\overrightarrow{k}),\quad\epsilon _{-1}(\overrightarrow{k})
=\epsilon (-\overrightarrow{k}),\quad\epsilon (
\overrightarrow{k})=\frac{1}{N}\sum\limits_{\overrightarrow{x}}t(
\overrightarrow{x})e^{i\overrightarrow{k}\overrightarrow{x}},\quad
\sum\limits_{\overrightarrow{k}}\epsilon (\overrightarrow{k})
=0,\quad G_{\sigma \sigma ^{\prime }}^{-1,-1}(k)=-G_{\sigma ^{\prime
}\sigma }^{1,1}(-k). \label{A5}
\end{eqnarray}
\end{widetext}%
%

Assuming that the system is in a paramagnetic state, that superconductivity
has a singlet character and using the definitions (8) together with the
additional ones: %
\begin{eqnarray}
\Lambda _{\sigma \sigma }^{1,1}(k) &=&\Lambda _{\sigma }(k),\Lambda _{\sigma
\overline{\sigma }}^{1,-1}(k)=Y_{\sigma \overline{\sigma }}(k),  \nonumber \\
\Lambda _{\overline{\sigma }\sigma }^{-1,1}(k)
&=&\overline{Y}_{\overline{ \sigma }\sigma }(k),  \label{A6}
\end{eqnarray}%
%
we obtain the following results:
%
\begin{widetext}
\begin{eqnarray}
G_{\sigma }(k) =\frac{\Lambda _{\sigma }(k)(1-\epsilon
(-\overrightarrow{k} )\Lambda _{\overline{\sigma }}(-k))-\epsilon
(-\overrightarrow{k})Y_{\sigma \overline{\sigma
}}(k)\overline{Y}_{\overline{\sigma }\sigma }(k)}{d_{\sigma
}(k)},\quad F_{\sigma \overline{\sigma }}(k) =\frac{Y_{\sigma
\overline{\sigma }}(k)}{ d_{\sigma }(k)},\quad
\overline{F}_{\overline{\sigma }\sigma }(k)=\frac{
\overline{Y}_{\overline{\sigma }\sigma }(k)}{d_{\sigma }(k)}, \\
d_{\sigma }(k) =(1-\epsilon (\overrightarrow{k})\Lambda _{\sigma
}(k))(1-\epsilon (-\overrightarrow{k})\Lambda _{\overline{\sigma
}}(-k)) +\epsilon (\overrightarrow{k})\epsilon
(-\overrightarrow{k})Y_{\sigma \overline{\sigma
}}(k)\overline{Y}_{\overline{\sigma }\sigma }(k),  \nonumber
\label{A7}
\end{eqnarray}
\end{widetext}%
%
which coincide with those found in the papers$^{[9,10]}$.

In spinor representation the system of equations \quad (A1-A7) has the form
%
\begin{eqnarray}
%
\widehat{G}=\widehat{\Lambda}+\widehat{\Lambda}\widehat{\epsilon}
\widehat{G},\label{A8}
\end{eqnarray}
%
were
%
\begin{eqnarray}
\widehat{G}= \left(
  \begin{array}{cc}
    G_{\sigma}(k) & F_{\sigma\overline{\sigma}}(k) \\
    \overline{F}_{\overline{\sigma}\sigma}(k) & -G_{\overline{\sigma}}(-k) \\
  \end{array}
\right),\nonumber
\end{eqnarray}
%
%
\begin{eqnarray}
\widehat{\Lambda}= \left(
  \begin{array}{cc}
    \Lambda_{\sigma}(k) & Y_{\sigma\overline{\sigma}}(k) \\
    \overline{Y}_{\overline{\sigma}\sigma}(k) & -\Lambda_{\overline{\sigma}}(-k) \\
  \end{array}
\right), \widehat{\epsilon}= \left(
  \begin{array}{cc}
    \epsilon(k) & 0 \\
    0 & -\epsilon(-k)\\
  \end{array}
\right)
\end{eqnarray}
%
By using equation (9) we can obtain
%
\begin{eqnarray}
%
T^{1,1}_{\sigma}(k)&=&\frac{\epsilon(\overrightarrow{k})(1-\epsilon(-\overrightarrow{k})\Lambda_{\overline{\sigma}}(-k))}{d_{\sigma}(k)},\nonumber\\
T^{1,-1}_{\sigma\overline{\sigma}}(k)&=&-\epsilon(\overrightarrow{k})\epsilon(-\overrightarrow{k})F_{\sigma\overline{\sigma}}(k),\\
T^{-1,1}_{\overline{\sigma}\sigma}(k)&=&-\epsilon(\overrightarrow{k})\epsilon(-\overrightarrow{k})\overline{F}_{\overline{\sigma}\sigma}(k).\nonumber\label{A10}
\end{eqnarray}
%
\end{document}